\documentclass{PoS}

\title{Microscopic Nuclear Structure and Reaction Calculations in the FMD Approach}

\ShortTitle{Microscopic Nuclear Structure and Reaction Calculations in the FMD Approach}

\author{\speaker{Thomas Neff}, Hans Feldmeier and Karlheinz Langanke\\
        GSI Helmholtzzentrum f{\"u}r Schwerionenforschung GmbH, Planckstra{\ss}e 1. 64291 Darmstadt, Germany\\
        E-mail: \email{t.neff@gsi.de}}

\abstract{
Fermionic Molecular Dynamics (FMD) is a microscopic many-body approach which has been used successfully to study the structure of light nuclei in the $p$- and $sd$-shell. FMD uses a Gaussian wave packet basis which contains the harmonic oscillator shell model and Brink-type cluster model wave functions as limiting cases. A realistic effective interaction that has been derived from the Argonne~V18 interaction by treating explicitly short-range central and tensor correlations is employed.

We present here a first application of the FMD approach to low-energy nuclear reactions, namely the $^3$He($\alpha$,$\gamma$)$^7$Be radiative capture reaction. We divide the Hilbert space into an external region where the system is described as $^3$He and $^4$He clusters interacting only via the Coulomb interaction and an internal region where the nuclear interaction will polarize the clusters. Polarized configurations are obtained by a variation after parity and angular momentum projection procedure with respect to the parameters of all single particle states. A constraint on the radius of the intrinsic many-body state is employed to obtain polarized clusters at desired distances. The boundary conditions for bound and scattering states are implemented using the Bloch operator.

The FMD calculations reproduce the correct energy for the centroid of the $3/2^-$ and $1/2^-$ bound states in $^7$Be. The charge radius of the ground state is in good agreement with recent experimental results. The FMD calculations also describe well the experimental phase shift data in the $1/2^+$, $3/2^+$ and $5/2^+$ channels that are important for the capture reaction at low energies. Using the bound and scattering many-body wave functions we calculate the radiative capture cross section. The calculated $S$ factor agrees very well, both in absolute normalization and energy dependence, with the recent experimental data from the Weizmann, LUNA, Seattle and ERNA experiments.
}

\FullConference{11th Symposium on Nuclei in the Cosmos\\
                 19-23 July 2010 \\
                 Heidelberg, Germany.}

\begin{document}

\newcommand{\nuc}[2]{$^{#2}${#1}}
\newcommand{\heag}{\nuc{He}{3}($\alpha$,$\gamma$)\nuc{Be}{7}}

\newcommand{\op}[1]{#1}
\newcommand{\ket}[1]{\big| {#1} \big> }
\newcommand{\braket}[2]{\big< {#1} \big| {#2} \big> }

\section{Introduction}

Low-energy nuclear reactions play an important role in many
astrophysical scenarios. In many cases experimental data are not
available at the energies relevant for the astrophysical
processes. Typically such reactions are described with potential models
which describe the reaction partners as point-like nuclei interacting
via nucleus-nucleus potentials fitted to experimental data on bound
and scattering states. In a microscopic \emph{ab initio} picture however, the system is
described as a many-body system of interacting nucleons. The wave
functions are fully antisymmetrized and realistic nucleon-nucleon
interactions are used.

In the Fermionic Molecular Dynamics (FMD) approach \cite{fmd08} we aim at a
consistent description of bound states, resonances and scattering
states using realistic low-momentum nucleon-nucleon
interactions obtained in the Unitary Correlation Operator Method (UCOM) \cite{ucom10}. Intrinsic many-body basis states are Slater determinants
using Gaussian wave packets as single-particle states. This basis
contains harmonic oscillator shell model and Brink-type cluster wave
functions as special cases. The symmetries of the system are restored
by projection on parity, angular momentum and total linear
momentum. The many-body eigenstates of the realistic Hamiltonian are obtained in multiconfiguration
mixing calculations. FMD has already been used successfully to describe nuclei in the $p$- and $sd$-shell, like the Hoyle state in \nuc{C}{12} \cite{hoyle07} or halo- and cluster-structures in Neon isotopes \cite{geithner08}.

In this contribution we present results for the \heag\/ capture cross section using the FMD approach. This reaction has been studied using potential models \cite{kim81,mohr09} and microscopic cluster models \cite{langanke86,kajino86} assuming \nuc{He}{3}+\nuc{He}{4} cluster wave functions. Polarization effects were considered by including the \nuc{Li}{6}+$p$ channel in \cite{mertelmeier86,csoto00}. Consistent \emph{ab initio} calculations starting from realistic interactions have not been possible up to now. First attempts using Variational Monte-Carlo \cite{nollett01} and the No-Core Shell Model \cite{navratil07b} only calculated the asymptotic normalization coefficients from the bound state wave functions.

\begin{figure}[b]
	\centering
  \includegraphics[width=0.5\textwidth]{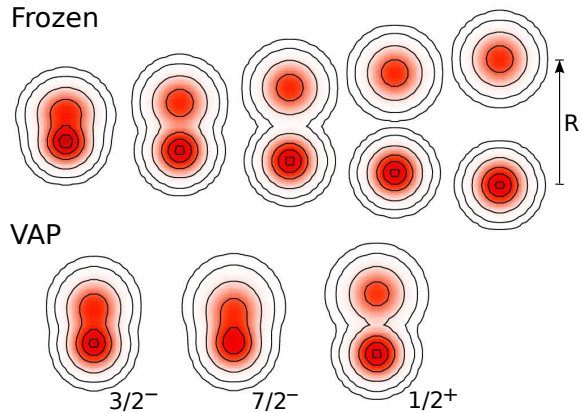}
	\caption{Density distributions of the intrinsic basis states. Top: frozen cluster configurations, bottom: selected polarized configurations obtained by variation after angular momentum projection.}
  \label{fig:intrinsic}
\end{figure}

\section{Fermionic Molecular Dynamics}

In FMD the intrinsic many-body basis states are Slater determinants
\begin{equation}
  \label{eq:fmdslaterdet}
  \ket{Q} = \op{\mathcal{A}} \bigl\{ \ket{q_1} \otimes \ldots \otimes
    \ket{q_A} \bigr\} \: ,
\end{equation}
using Gaussian wave packets as single-particle states
\begin{equation}
  \braket{\vec{x}}{q_k} =
  \exp \biggl\{ -\frac{(\vec{x} -\vec{b}_k)^2}{2 a_k} \biggr\} \otimes
  \ket{\chi^\uparrow_{k},\chi^\downarrow_{k}} \otimes \ket{\xi_k} \: .
\end{equation}
The complex parameters $\vec{b}_k$ encode the mean positions and momenta of the wave packets, the width parameters $a_k$ are variational, the spins can assume any direction. To restore the symmetries of the Hamiltonian the intrinsic wave function is projected on parity, angular momentum and total linear momentum $\vec{P}=0$.
\begin{equation}
  \ket{Q; J^\pi MK; \vec{P}=0} =
  \op{P}^{J}_{MK} \op{P}^\pi \op{P}^{\vec{P}=0}\ket{Q} \: .
\end{equation}

As in a microscopic cluster model we have ``frozen'' cluster configurations where \nuc{He}{4} and \nuc{He}{3} clusters are located at a distance $R$. These Slater determinants can be rewritten using RGM basis states \cite{baye77,descouvemont10}. The RGM representation can be used to match the logarithmic derivative of the relative wave function of the clusters to the asymptotic Whittaker or Coulomb wave functions. In the interaction region we have additional FMD basis states that are necessary to describe the polarization of the clusters. These basis states are generated by variation of all single-particle parameters after parity and angular momentum projection. We use a constraint on the radius of the intrinsic states to generate configurations corresponding to different cluster distances. In Fig.~\ref{fig:intrinsic} we show intrinsic density distributions for some frozen and polarized basis states.

\section{Bound and Scattering States}

\begin{figure}[b]
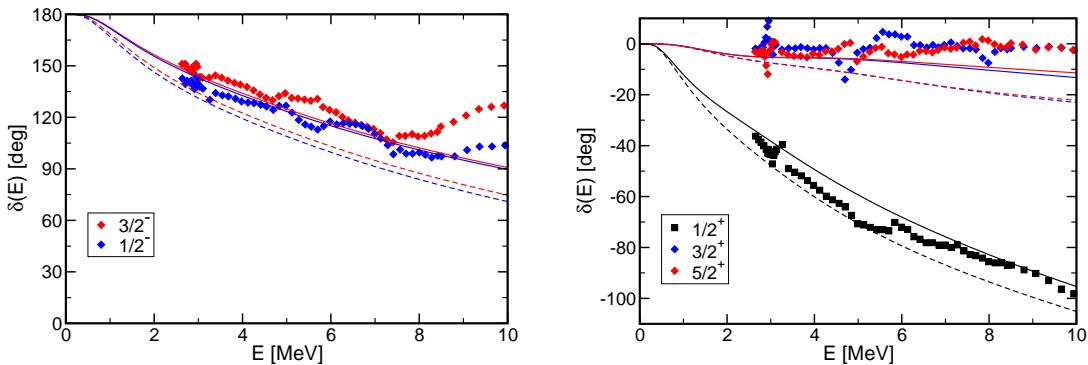

  \includegraphics[width=0.45\textwidth]{figs/FMD-He4-He3-Phaseshifts-31-.eps}\hfil
	\includegraphics[width=0.45\textwidth]{figs/FMD-He4-He3-Phaseshifts-135+.eps}
	\caption{\nuc{He}{4}-\nuc{He}{3} scattering phase shifts for $P$-waves left and the $S$ and $D$-waves right. Dashed lines show results using only frozen cluster configurations, solid lines show results for the full model space.}
	\label{fig:phaseshifts}
\end{figure}

Using the microscopic $R$-matrix approach of the Brussels group \cite{baye77,descouvemont10} we solve the Schr{\"o}dinger equation with the proper boundary conditions for bound and scattering states. With frozen configurations alone the $3/2^-$ and $1/2^-$ states are bound by only 240~keV and 10~keV, respectively. In the full Hilbert space, including the polarized configurations, we obtain binding energies of 1.49~MeV and 1.31~MeV with respect to the cluster threshold. We therefore reproduce the centroid energy but underestimate the splitting of the $3/2^-$ and $1/2^-$ states. This will effect the branching ratio but not the total cross section which essentially only depends on the centroid energy. The calculated charge radius of 2.67~fm is in good agreement with the experimental value of 2.647(17)~fm \cite{noertershaeuser09}. In Fig.~\ref{fig:phaseshifts} we present the calculated phase shifts together with the experimental data \cite{spiger67}. Again we see a sizable effect when we compare the results using only the frozen configurations with the results obtained in the full Hilbert space including polarization effects.

\section{Capture Cross Section}

\begin{figure}[t]
	\centering
  \includegraphics[width=0.6\textwidth]{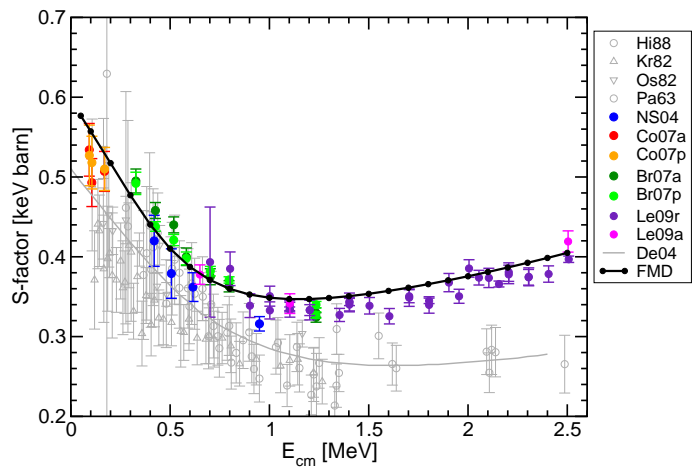}
	\caption{The astrophysical $S$-factor for the \heag\/ reaction. The FMD result is given by a black solid line. Old data are shown as gray symbols together with an $R$-matrix fit. Recent data are shown by colored symbols.}
	\label{fig:sfactor}
\end{figure}

Using the microscopic bound and scattering wave functions we calculate the radiative capture cross section. We have contributions from the internal region calculated with the FMD many-body wave functions and from the external region up to large cluster distances calculated with the matched Whittaker and Coulomb solutions. At low energies we can restrict ourselves to electric dipole transitions from the $S$- and $D$-wave channels. The obtained cross section, presented in form of the astrophysical $S$ factor is shown in Fig.~\ref{fig:sfactor}. It agrees very well both in absolute normalization and energy dependence with the recent experimental results obtained at the Weizmann Institute \cite{narasingh04}, the LUNA collaboration \cite{confortola07}, at Seattle \cite{brown07} and by the ERNA collaboration \cite{dileva09}.

\begin{thebibliography}{10}

\bibitem{fmd08}
T.~Neff and H.~Feldmeier, {\it Clustering and other exotic phenomena in
  nuclei},  {\em Eur. Phys. J Special Topics} {\bf 156} (2008) 69.

\bibitem{ucom10}
R.~Roth, T.~Neff, and H.~Feldmeier, {\it Nuclear structure in the framework of
  the {U}nitary {C}orrelation {O}perator {M}ethod},  {\em Prog. Part. Nucl.
  Phys.} {\bf 65} (2010) 50.

\bibitem{hoyle07}
M.~Chernykh, H.~Feldmeier, T.~Neff, P.~von Neumann-Cosel, and A.~Richter, {\it
  On the structure of the {H}oyle {S}tate in ${}^{12}${C}},  {\em Phys. Rev.
  Lett.} {\bf 98} (2007) 032501.

\bibitem{geithner08}
W.~Geithner, T.~Neff {\it et al.}, {\it Masses and
  {C}harge {R}adii of $^{17-22}${N}e and the {T}wo-{P}roton-{H}alo {C}andidate
  $^{17}${N}e},  {\em Phys. Rev. Lett.} {\bf 101} (2008) 252502.

\bibitem{kim81}
B.~T. Kim, T.~Izumoto, and K.~Nagatani, {\it Radiative capture reaction
  $^3${H}e($\alpha$,$\gamma$)$^7${B}e at low energies},  {\em Phys. Rev. C}
  {\bf 23} (1981) 33.

\bibitem{mohr09}
P.~Mohr, {\it Low-energy $^3${H}e($\alpha$,$\alpha$)$^3${H}e elastic scattering
  and the $^3${H}e($\alpha$,$\gamma$)$^7${B}e reaction},  {\em Phys. Rev. C}
  {\bf 79} (2009) 065804.

\bibitem{langanke86}
K.~Langanke, {\it Microscopic potential model studies of light nuclear capture
  reactions},  {\em Nucl. Phys.} {\bf A457} (1986) 351.

\bibitem{kajino86}
T.~Kajino, {\it The $^3${H}e($\alpha$,$\gamma$)$^7${B}e and
  $^3${H}($\alpha$,$\gamma$)$^7${L}i {R}eactions at {A}strophysical
  {E}nergies},  {\em Nucl. Phys.} {\bf A460} (1986) 559.

\bibitem{mertelmeier86}
T.~Mertelmeier and H.~M. Hofmann, {\it Consistent cluster model description of
  the electromagnetic properties of {L}ithium and {B}eryllium nuclei},  {\em
  Nucl. Phys.} {\bf A459} (1986) 387.

\bibitem{csoto00}
A.~Cs{\'o}to and K.~Langanke, {\it Study of the
  $^3${H}e($^4${H}e,$\gamma$)$^7${B}e and $^3${H}($^4${H}e,$\gamma$)$^7${L}i
  {R}eactions in an {E}xtended {T}wo-{C}luster {M}odel},  {\em Few-Body
  Systems} {\bf 29} (2000) 121.

\bibitem{nollett01}
K.~M. Nollett, {\it Radiative $\alpha$-capture cross sections from realistic
  nucleon-nucleon interactions and variational {M}onte {C}arlo wave functions},
   {\em Phys. Rev.} {\bf C63} (2001) 054002.

\bibitem{navratil07b}
P.~Navr{\'a}til, C.~A. Bertulani, and E.~Caurier, {\it
  $^7${B}e(p,$\gamma$)$^8${B} {S}-factor from \emph{ab initio} wave functions},
   {\em Nucl. Phys.} {\bf A787} (2007) 539c.

\bibitem{baye77}
D.~Baye, P.-H. Heenen, and M.~Libert-Heinemann, {\it Microscopic {R}-{M}atrix
  {T}heory in a {G}enerator {C}oordinate {B}asis},  {\em Nucl. Phys.} {\bf
  A291} (1977) 230.

\bibitem{descouvemont10}
P.~Descouvemont and D.~Baye, {\it The ${R}$-matrix theory},  {\em Rep. Prog.
  Phys.} {\bf 73} (2010) 036301.

\bibitem{noertershaeuser09}
W.~N{\"o}rtersh{\"a}user {\it et al.}, {\it Nuclear {C}harge {R}adii
  of $^{7,9,10}${B}e and the {O}ne-{N}eutron {H}alo {N}ucleus $^{11}${B}e},
  {\em Phys. Rev. Lett} {\bf 102} (2009) 062503.

\bibitem{spiger67}
R.~J. Spiger and T.~A. Tombrello, {\it Scattering of {H}e$^3$ by {H}e$^4$ and
  of {H}e$^4$ by {T}ritium},  {\em Phys. Rev.} {\bf 163} (1967) 964.

\bibitem{narasingh04}
B.~S. Nara~Singh, M.~Hass, Y.~Nir-El, and G.~Haquin, {\it New {P}recision
  {M}easurement of the $^3${H}e($^4${H}e,$\gamma$)$^7${B}e {C}ross {S}ection},
  {\em Phys. Rev. Lett.} {\bf 93} (2004) 262503.

\bibitem{confortola07}
F.~Confortola {\it et al.}, {\it
  Astrophysical ${S}$ factor of the $^3${H}e($\alpha$,$\gamma$)$^7${B}e
  reaction measured at low energy via detection of prompt and delayed $\gamma$
  rays},  {\em Phys. Rev. C} {\bf 75} (2007) 065803.

\bibitem{brown07}
T.~A.~D. Brown, C.~Bordeanu, K.~A. Snover, D.~W. Storm, D.~Melconian, A.~L.
  Sallaska, S.~K.~L. Sjue, and S.~Triambak, {\it $^3${H}e + $^4${H}e
  $\rightarrow$ $^7${B}e astrophysical ${S}$ factor},  {\em Phys. Rev. C} {\bf
  76} (2007) 055801.

\bibitem{dileva09}
A.~Di~Leva {\it et al.}, {\it Stellar and {P}rimordial {N}ucleosynthesis
  of $^7${B}e: {M}easurement of $^3${H}e($\alpha$,$\gamma$)$^7${B}e},  {\em
  Phys. Rev. Lett.} {\bf 102} (2009) 232502.

\end{thebibliography}
\providecommand{\href}[2]{#2}\begingroup\raggedright\endgroup

\end{document}